\documentstyle[aas2pp4,flushrt,psfig2]{article}

\newcommand{\bm}[1]{\mbox{\boldmath $#1$}}

\newcommand{\be}{\begin{equation}}
\newcommand{\ee}{\end{equation}}
\newcommand{\bea}{\begin{eqnarray}}
\newcommand{\eea}{\end{eqnarray}}

\newcommand{\simless}{\mathbin{\lower 3pt\hbox
 {$\rlap{\raise 5pt\hbox{$\char'074$}}\mathchar"7218$}}}

\newcommand{\simgreat}{\mathbin{\lower 3pt\hbox
 {$\rlap{\raise 5pt\hbox{$\char'076$}}\mathchar"7218$}}}

\begin{document}

\title{Interstellar Scintillation of Pulsar B0809+74}
\author{B. J. Rickett\altaffilmark{1}, Wm. A. Coles\altaffilmark{2}}
\affil{Department of Electrical and Computer Engineering,
University of California, San Diego, CA 92093}
\author{Jussi Markkanen\altaffilmark{3}}
\affil{EISCAT-Geophysical Observatory, SF-99600 Sodankyl\"{a},  Finland}

\altaffiltext{1}{e-mail: rickett@ece.ucsd.edu}
\altaffiltext{2}{e-mail: coles@ece.ucsd.edu}
\altaffiltext{3}{e-mail: jussi@eiscat.sgo.fi}

\begin{abstract}
Weak interstellar scintillations of pulsar B0809+74
were observed at two epochs using a 30m
EISCAT antenna at 933 MHz. These have been used to constrain the
spectrum, the distribution and the transverse velocity of 
the scattering plasma with respect to the local standard of rest (LSR).
The Kolmogorov power law is a satisfactory model for the 
electron density spectrum at scales between $2\times 10^7$m and $10^9$m.
We compare the observations with model calculations 
from weak scintillation theory and the known transverse velocities 
of the pulsar and the Earth.  The simplest model is that
the scattering is uniformly distributed along the 310 pc line of sight 
($l=140^\circ$, $b=32^\circ$)
and is stationary in the LSR.  With the scattering measure 
as the only free parameter, this model fits the data within the 
errors and a range of about $\pm10$ km~s$^{-1}$ in velocity is also allowed.  
The integrated level of turbulence is low, 
being comparable to that found toward PSR B0950+08,
and suggests a region of low local turbulence over
as much as 90$^{\circ}$ in longitude including the galactic anti-center.
If, on the other hand, the scattering occurs in
a compact region, the observed time
scales require a specific velocity-distance relation. 
In particular, enhanced scattering in a shell at the edge of the local 
bubble, proposed by Bhat et al.\ (1998),
near 72 pc toward the pulsar, must be moving at about
$\sim 17$ km~s$^{-1}$; 
however,  the low scattering measure argues against
a shell of enhanced scattering in this direction.
The analysis also excludes scattering in the termination
shock of the solar wind or in a nebula associated with the pulsar.
\end{abstract}


\section{Introduction}
\label{sec:intro}

The use of pulsar scintillation to probe the interstellar medium has
been complicated by the fact that most observations below about 1 GHz
are in the strong scintillation regime. In this regime there are both
refractive and diffractive components in the intensity scintillation
pattern, and neither spatial scale is known a priori. Furthermore the
contributions from various locations along the path of propagation do
not add in a simple linear fashion (see Rickett, 1990 and Narayan, 1992
for reviews).  Weak scintillation is much simpler to interpret since
contributions from various locations add linearly and the spatial 
scale of each contribution depends only weakly on unknown 
parameters such as the spectral exponent and the inner scale. 
The temporal scale of each contribution depends primarily on
its distance and its velocity. 
In principle, it is feasible to ``invert'' the observations to 
estimate the spatial spectrum of the electron density and, by 
exploiting the variable 
velocity of the Earth, to investigate
distribution of the scattering plasma and its velocity. 

Observations can be made in weak scintillation by using higher
frequencies and/or nearby pulsars. However it is difficult to obtain a
stable estimate of the covariance function because the time scale of
the scintillation is of the order of an hour. Thus a given observation
will typically contain only a few time scales. Earlier observations by
Backer (1975) at 3 GHz and Malofeev et al (1996) at 3, 5 and 8 GHz have
confirmed that weak scattering occurs more or less as expected, but
have not been long enough to provide the statistical accuracy necessary
for detailed model fitting.

In this paper we report two longer measurements of a circumpolar
pulsar in weak scintillation. The temporal statistics of these
data sets are adequate to 
estimate the spectrum over a decade in wavenumber and put constraints on
the distribution and velocity of scattering plasma.  
We have developed a procedure for the 
analysis that may be more generally useful.

\section{Observations and Data Analysis}
\label{sec:obs}

The pulsar B0809+74 was selected because earlier observations at
lower frequencies suggested that it should be in weak scintillation at
933 MHz, which is the only frequency available on the 30m EISCAT
antenna at Sodankyl\"{a}, Finland, and because it is circumpolar 
at that latitude (68N).
The receiver passband is 8 MHz centered on 933 MHz and one
linear polarization was used. The use of one linear
polarization introduces a potential bias in the
analysis since the pulsar is partially polarized. As a result
the apparent flux varies due to the rotation of the antenna
and the Faraday rotation in the ionosphere. This was thought
to be negligible at 933 MHz, but was later found to be
important. The effect was eliminated by selecting an
unpolarized portion of the pulse profile as discussed below.

Impulsive interference is common at Sodan\-kyl\"{a} and the square law
detector was designed to reduce the contribution from very short
noise pulses.  The pulsar intensity could be adequately sampled
at 100 samples per sec (sps), however it was actually sampled at 100 Ksps using a very
broad band detector.  Pulses exceeding about 10 standard deviations
were clipped at the sampler.  The 100 Ksps series was then smoothed and
decimated by a factor of 1000. This process reduced the contribution of
impulsive interference to a low level without distorting the pulsar
signal, and also reduced the rms
quantization noise by a factor of 30.  We
observed continuously for 75 hours starting at 20 hr UT on 1996
April 8, although the observations were stopped due to system errors and
restarted several times.  We also observed a second epoch from 23 hr UT
on 1998 September 30 for 85 hrs.

The data were summed (off-line) according to the apparent pulsar period
(Doppler-shifted to the observatory).  Each period ($\approx$1.29 s)
was tested for interference using the ratio of the rms to the mean
off-pulse power. The expected value of this ratio is about 0.003 and
its standard deviation for a single pulse period is about 0.0003.
Periods for which this ratio exceeded 0.005, about 7 standard
deviations, were rejected. Successive integrations of the pulse profile
were computed every 2 minutes. These were examined by hand and obvious
interference was removed. The automatic interference test was very
effective so less than 1\% of the 2 minute averages were edited
manually. Pulse profiles were also computed every 10 minutes.
Both 2 and 10 minute averages were used in subsequent work.
The average pulse profile was used to estimate the energy in an 
on-pulse window and also in an equal off-pulse window. First
an estimate of the background power was subtracted, then the
energy in each of the two windows was computed. 

\begin{figure}[tbh]
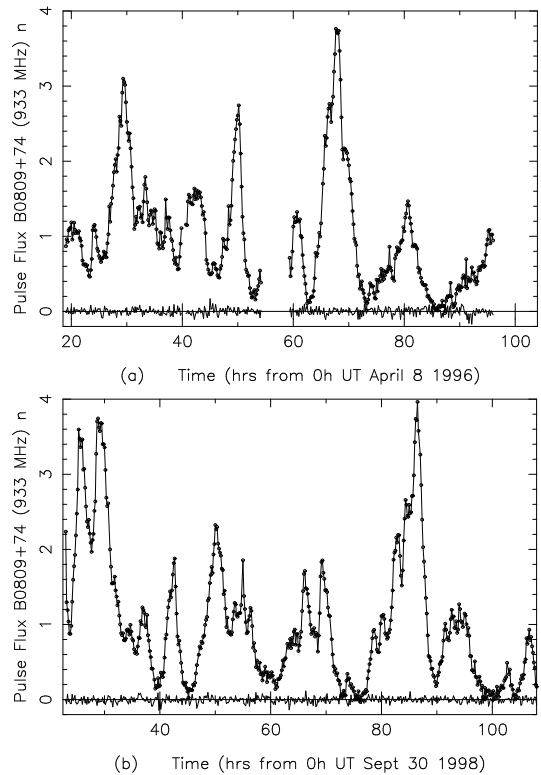

\begin{tabular}{c}
\psfig{figure=f1a.ps,width=7cm,rotate=90} \\
\psfig{figure=f1b.ps,width=7cm,rotate=90}
\end{tabular}
\caption{Observations of the total pulse flux of B0809+74 at 933 MHz,
normalized to the mean flux over the observing period. The samples
plotted are 10 minute averages. The lower traces are computed from an
off-pulse window to provide a noise estimate. (a) Data from April 1996.
(b) data from October 1998.} 
\label{fig:f1} 
\end{figure}

Initially we used an on-pulse window which included the entire 
pulse, however a referee, (Michael Kramer), pointed out that 
B0809+74 has 30\% linear polarization (Gould and Lyne, 1998). 
Examination of our pulse profiles revealed a 24 hr periodicity
in the leading edge, 
which could be attributed to the linear polarization of the pulsar, 
the rotation of the telescope, and the diurnal change in the Faraday 
rotation in the ionosphere. This contributed about 10\% of the
variance in the pulse flux. 
Although this is not large it was greater than
our estimation error and significantly 
altered our conclusions.
The Gould and Lyne observations show strong linear polarization in 
the leading half of the pulse with no rotation of position
angle, but the trailing half of the pulse 
is essentially unpolarized. Thus we changed our pulse energy
computation to use a 50 msec on-pulse window starting at the
pulse peak and including all the trailing half. With this window
the polarization modulation was not detectable. The resulting
time series of on-pulse power and off-pulse noise are displayed in Figure 
\ref{fig:f1} with 10 minute resolution. The flux density was not 
calibrated absolutely, it has been normalized to unity averaged 
over the entire observing period.

In the absence of explicit flux density calibration, we used the
variation of the system noise level to obtain a limit on the gain
variation. The off pulse system noise varied almost regularly with a
period of 24 hrs and an amplitude of about 6\% peak to peak.  We assume
that most of this variation is due to changes in the atmospheric and
ground noise contributions, as the zenith angle varied from 6$^\circ$
to 38$^\circ$ and the azimuth remained in the northern quadrant during
the 24 hr period. We therefore conclude that the gain variation is
periodic over  24 hr  with an amplitude considerably less than 6\% and
thus is negligible compared to the large variations over a few hours
observed in the pulse intensity.

The second order temporal statistics, i.e. the power spectrum
and the autocovariance function, can be theoretically modelled
and used to estimate the distribution of scattering material,
its velocity and some characteristics of the turbulent
spectrum. Although the two statistics are simply related by a
Fourier transform, their estimation errors are quite different
and both are useful for different purposes.

To estimate the time scale we computed the structure function rather
than the autocovariance function, since the structure function is a
more reliable estimator when the number of independent samples of the
scintillations is not very large, as is the case here.  For this
purpose we used the 10 min data because the noise correction for these
data is negligible and the resolution is more than adequate.  The
structure function was computed in the usual way from the series of
pulsar energies $I_j$ (normalized by their estimated mean) sampled at
times $t_j$.
\begin{equation} 
\hat D_{I}(\tau_{l}) =
{\Sigma^{i,j}_{l} [ I_{i}- I_{j}]^2 / \Sigma^{i,j}_{l} 1 }
\label{eq:dhat,disc} 
\end{equation}  
Here $\Sigma^{i,j}_{l}$ represents the summation over all data pairs 
with time difference $t_i - t_j$ lying within a window of width 
$\delta p$ centered on $\tau_{l} = l \delta p$, where $l$ is an 
integer and $\delta p$ is the resolution of the data. The maximum lag 
computed was three quarters of the total duration of the observations. 

\begin{figure}[bth]
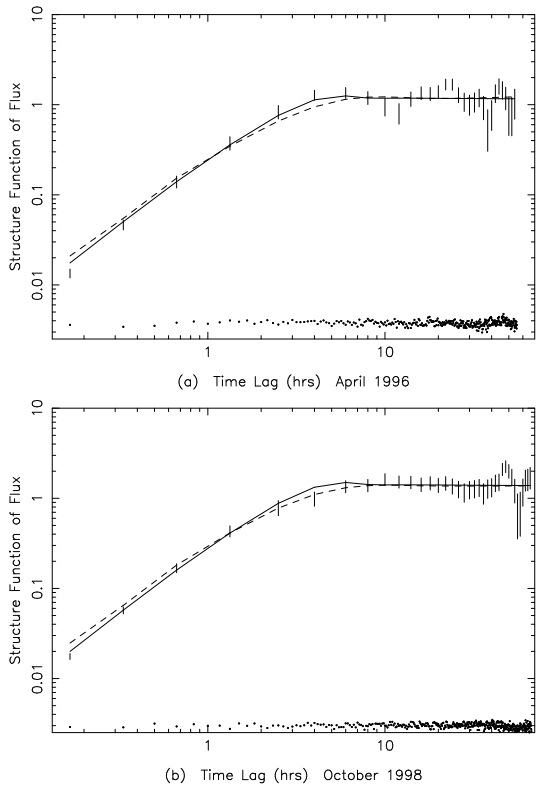

\begin{tabular}{c}
\psfig{figure=f2a.ps,width=7cm,rotate=90} \\
\psfig{figure=f2b.ps,width=7cm,rotate=90}
\end{tabular}
\caption{Structure functions of pulse flux time series shown
in Figure \ref{fig:f1} corrected for receiver noise and
intrinsic variation. The data are marked by $\pm \sigma$ error 
bars: (a) April 1996; (b) October 1998.
The solid lines are the best fit theoretical models for a thin 
screen. The dashed lines are best fit models for a uniform extended
scattering medium (as described in section 3.2).
The dots represent the structure function of the receiver noise.
}
\label{fig:lgstrfn}
\end{figure}

Errors in the structure function result from additive receiver 
(radiometer) noise, intrinsic pulse to pulse variation, and from 
estimation errors due to the finite observing span. The receiver 
noise and intrinsic variation are effectively white and
independent of the scintillation. They represent additive
constants in the structure function which can be subtracted. 
The receiver noise contribution can be estimated very accurately 
from the off-pulse noise. To measure the intrinsic variation we 
used the 2 min data. The first point in the structure function 
of the 2 min data is dominated by intrinsic variation and 
receiver noise. This was used to estimate the contribution of 
intrinsic variation to the structure function of the 10 min 
data with very good accuracy. The corrected structure function
of the 10 min data is shown in Figure \ref{fig:lgstrfn}.

The estimation errors in the structure function due to the finite 
data length are substantial; they are derived in Appendix B.
The $\pm\sigma$ estimation errors, given by equation (\ref{eq:drms}),
are indicated by the error bars in Figure \ref{fig:lgstrfn}.
As discussed in the Appendix the fractional error due to estimation
error decreases toward small time lags but is partially compensated by
the addition of the errors due to noise.  The estimation errors are
correlated over about an octave in time lag below saturation. At
larger lags they are correlated over a characteristic time scale
defined as the 50\% width of the autocovariance.  For model fitting
purposes, the lags were decimated to reflect this distribution
of independent estimates.

The power spectra were computed from the 2 min data using a
direct Fourier transform procedure. The 1996 data was divided
at the data gap and the two halves were transformed separately
and their spectra added. The 1998 data was transformed in one
block. The spectra were computed with a raw resolution of 
7.3$\times 10^{-3}$ cycles per hour (cph) 
and normalized so that the integral
under the spectrum is the variance. They were boxcar smoothed 
and decimated by a factor which increased from 3
for the lowest 4 points to 125 for the highest frequencies.
This reduces the estimation error and broadens the resolution. 
The white noise contribution from receiver noise and instrinsic 
variation was estimated from the frequencies above 5 cph,
where the scintillation power is negligible, and subtracted
from the spectral estimate. The final power spectra are
plotted in Figure~3.
The vertical bars indicate the $\pm\sigma$ estimate error
and the horizontal bars indicate the frequency resolution.
Power spectra are convenient for model fitting, particularly 
at the high frequencies, because the estimation errors are 
independent and their variance is known. Thus it is easy to 
obtain reliable confidence intervals.

\begin{figure}[bth]
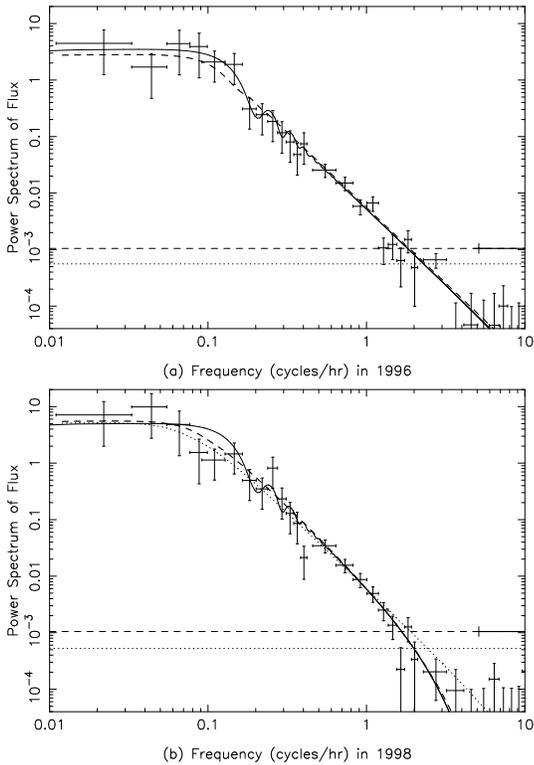

\begin{tabular}{c}
\psfig{figure=f3a.ps,width=7cm,rotate=90} \\
\psfig{figure=f3b.ps,width=7cm,rotate=90}
\end{tabular}
\caption{Power spectra of the pulse flux time series shown
in Figure \ref{fig:f1} corrected for receiver noise and
intrinsic variation, (a) 1996 and (b) 1998.
The data are marked by $\pm \sigma$ error bars and the
resolution is indicated by the horizontal bar.
The off-pulse noise spectrum is marked by a dotted horizonal
line. The sum of the off-pulse noise and the intrinsic noise
is marked by a dashed horizontal line. The frequency range
from which the intrinsic contribution was estimated is marked
with a heavy horizontal line.
The best fit theoretical models for a thin screen are shown by 
solid lines. 
The best fit theoretical models for a uniform medium which is 
stationary in the LSR are shown by dashed lines.
The best fit models in 1996 have zero inner scale, but in 1998
they have $s_i = 2 \times 10^7$ m.
A best-fit theoretical model for a uniform medium with zero
inner scale is also drawn over the 1998 data as a
dotted line for comparison.
}
\label{fig:pspec}
\end{figure}

Two basic parameters which describe the flux time series are the
normalized standard deviation 
or scintillation index $m_{\rm w}$, and the 
time scale $\tau_{\rm w}$ at which the correlation function falls
to 50\%. They were derived from the structure function using
the fact that $D_I(\tau)$ saturates at $2 m_{\rm w}^2$ 
for large time lags and estimating the time scale
as the lag where $D_I(\tau)$ crosses half the saturation
level (i.e. $D_I(\tau_w)=m_w^2$).
By this method we found $m_{\rm w} = 0.77 \pm 0.2$ and
$\tau_{\rm w} = 1.93 \pm 0.2$ hr for the 1996 observations and
$m_{\rm w} = 0.84 \pm 0.15$, and
$\tau_{\rm w} = 2.19 \pm 0.2$ hr for the 1998 observations.

\section{Model Fitting and Interpretation}
\label{sec:mod}
Even though the observed scintillation indices ($\approx 0.8$)
are not much less than unity,  we initially interpreted 
the results using weak scintillation theory.
The advantage of weak scintillation theory is that the observed
structure functions and power spectra are linear integrals 
contributed from all points on the line of sight as described
by the Born approximation discussed in Appendix A. 
Under weak scintillation we can do quantitative modelling of the 
density spectrum and the velocity distribution of the scattering 
medium. However when the scintillation index is approaches unity, 
there can be significant departures from the linear Born model. 
Simulations by Frehlich (1999, private communication)
corresponding to our observed $m_w \approx 0.8$ show that the 
primary effect of incipient strong scattering is to broaden 
the spectrum by a factor of 1.20$\pm$0.02 and to reduce the 
time scale $\tau_w$ by the same factor. The scintillation
index $m_w$ is also reduced by 4\% from the weak scattering
approximation. 
The statistical error in $m_w$ is about 20\% so the 4\% bias 
due to incipient strong scintillation is negligible. 
However the statistical error in $\tau_w$ is only 10\%, so we 
have divided all our Born model calculations by the factor of
1.20 to ensure that the bias is less than the statistical error. 
Strong scattering also smoothes the ``Fresnel ripples'' in the 
spectrum and rounds the ``knee'' of the structure function, but 
this is a second order effect.

We assume that the density spectrum is an isotropic
Kolmogorov power law, and also consider the effect of an 
inner scale. We need only consider one basic geometry: the pulsar is a point 
source at a distance L from the Earth and a thin scattering screen 
is located $z_p$ from the pulsar and $z_o$ from the earth, where 
$z_p + z_o = L$. To compute the temporal statistics we also need 
the velocity of the Earth, the pulsar, and the scattering screen.
We can then compute the effect of a distribution of scattering
material along the line of sight by integrating over all such
elementary screens. 

We have referred all velocities to the ``local standard of rest''
(LSR), since that appears to be the logical reference frame for the
velocity of the interstellar plasma between the Earth and the pulsar.
In this frame the solar velocity is 20 km~s$^{-1}$ directed towards
right ascension of 18 hrs and declination of +30$^\circ$.
The pulsar velocity 
($V_{\alpha} = 31 \pm 10$ km~s$^{-1}$; $V_{\delta} = -55 \pm 9$ km~s$^{-1}$ in this frame)
is determined from its proper motion (Lyne, Anderson, and Salter, 
1982) and its distance (310 pc) from the
model of Taylor and Cordes, 1993 [TC93]. 

In the following we will compare the observations with two models: 
(1) a single thin screen and (2) a uniform distribution. 
We find satisfactory agreement for both models, and
then explore what range of LSR
velocities is allowed for the scattering medium.

\subsection{Thin Screen Model}

The intensity structure function for weak scattering by a thin 
screen with a simple Kolmogorov spectrum (equation \ref{eq:kolmo}), 
has the universal form (\ref{eq:Ddscreen}).
The time scale $\tau_{\rm w} =$ 0.98 $r_f / V_{\rm eff}$, where
$r_f = \sqrt{z_o z_p/L k}$ is the ``Fresnel scale'', i.e. the 
spatial scale at the screen that contributes most of the 
scintillation, and $V_{\rm eff}$ is the velocity of the scattering 
plasma across the line of sight given by equation (\ref{eq:veff}).
(It should be noted that the spatial scale at the observing
plane is expanded by spherical divergence to $r_f L / z_p$).

\begin{figure}[bth]
\psfig{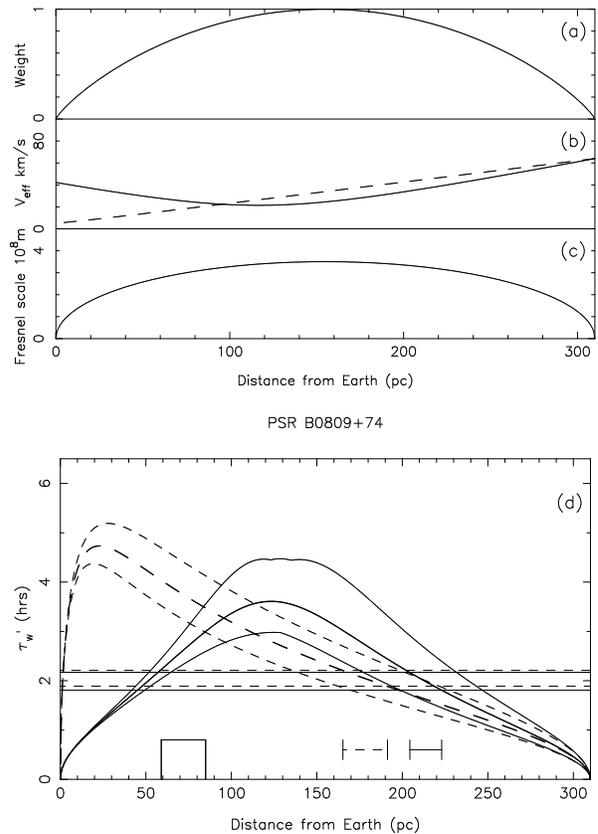}
\caption{Dependence of weak scintillation parameters
on screen distance: (a) contribution to variance;
(b) effective velocity $V_{\rm eff}$ with respect to the LSR
(solid line 1996, dashed 1998); (c) Fresnel scale;  and
(d) time scale (solid line 1996, dashed 1998). 
The time scales have been reduced by a factor 1.20 from the Born
theory to account for incipient strong scattering.
The thin lines show the envelope of time
scales corresponding to errors in the pulsar proper motion.
The confidence intervals for the observed time scale are shown
as horizonal line pairs near 2 hrs (solid line 1996, dashed 1998). 
The bars indicate allowed distances to the 
screen at each epoch assuming the nominal pulsar velocity.
The rectangle indicates the distance to the shell of enhanced 
scattering proposed by Bhat et al. (1998). }
\label{fig:tweak1}
\end{figure}

The best fit thin screen models are drawn over the observed
structure functions as solid lines on Figure \ref{fig:lgstrfn}.  
The structure function at the smallest lag was excluded
from the fit, in order to avoid the possible influence of an 
inner scale and because the correction for intrinsic pulse 
variation has uncertainties not reflected in the error bars 
at the smallest lag. 
As discussed earlier the lag values
were chosen to sample approximately independent estimates
and two parameters were fitted.  This gave
approximately 28 and 34 degrees of freedom,
respectively in 1996 and 1998.
The minimum $\chi^2$ were  34 and 30, respectively,
which indicates satisfactory fits to the data; the differences 
reflect the fact that the errors in the saturated portion of the
structure functions are noticeably different in the two years.
We use these fitted models to make refined estimates of the
time scale and its errors: 
$\tau_{\rm w} = 1.99 \pm 0.18$ hr and
$\tau_{\rm w} = 2.05 \pm 0.16$ hr for
the 1996 and 1998 observations, respectively.
These time scale estimates average 
over a range of time lags and so reduce the effect of the 
particular errors near where the structure function crosses 
half its saturation level. Hence we also use these same time
scale estimates to characterize the data in section 3.2 
on the extended medium model.

Weak scintillation theory predicts the relative contribution 
to the variance $m_{\rm w}^2$ and the Fresnel scale $r_f$ 
for a screen at position $0 < z_o < L$, which are plotted 
in panels (a) and (c) of Figure \ref{fig:tweak1}.
The effective velocity $V_{\rm eff}$ for the mean time of the 1996 
observations is plotted as a solid line and that for the 
1998 observations is plotted as a dashed line in panel (b)
under the assumption that the scattering plasma is stationary
with respect to the LSR.
The predicted time scales $\tau_{\rm w}' = 0.82\; r_f / V_{\rm eff}$
(corrected for $m_w = 0.8$)
are plotted in the lower panel Figure \ref{fig:tweak1}(d) as a thick
solid line for 1996 and a thick dashed line for 1998.
The uncertainty limits in these time scales caused by errors in the
pulsar proper motion are indicated by light solid and dashed
lines at $\pm\sigma$.
Also shown are the confidence intervals for the measured time scales 
as horizontal pairs of lines.

\begin{figure}[bth]
\psfig{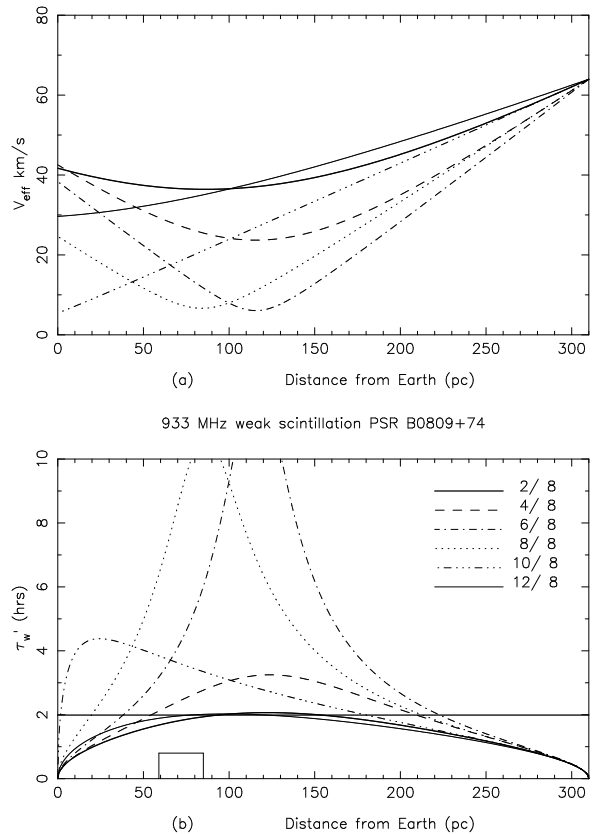}
\caption{Dependence of weak scintillation parameters on screen 
distance plotted for six epochs as indicated by month/day:
(a) $V_{\rm eff}$ with respect to the LSR;
(b) Scintillation time scale (with 20\% reduction to correct for 
incipient strong scattering). }
\label{fig:tweak6}
\end{figure}

Neglecting the error in the proper motion for the moment,
one can see that there are two locations at which a thin
screen would provide the time scale measured in 1996, 
at $51\pm6$ pc and $213\pm9$ pc. 
After analyzing the 1996 observations we 
realized that this ambiguity could be resolved with additional
observations at a different epoch (we thank Don Backer
for pointing this out). Then the
Earth's velocity would be different and so curves for
$V_{\rm eff}(z_o)$ and $\tau_w(z_o)$ would also be different.
This is shown in Figure \ref{fig:tweak6} where
we plot $V_{\rm eff}(z_o)$
and $\tau_{\rm w}(z_o)$ for six observing epoch's spaced
every two months on the 8th of each numbered month.
One can see that the velocity near the
pulsar does not change much but the velocity near the Earth
changes dramatically. If the scattering screen were
located near 213 pc, then the time scale would remain near 2
hrs. However if the scattering screen were located near 51 pc,
the time scale would be much larger if observations were made 
between July and October. Accordingly we chose to reobserve 
the source in October 1998.  One can see from Figure 
\ref{fig:tweak1}(d) that the 1998 observations 
are consistent with distances of 1$\pm$0.1 pc or 178$\pm$12 pc.  
The ambiguity is thus resolved in favor of the larger distance.  
However the match is not perfect as the $\pm\sigma$ error bars
do not overlap. The probability of this occuring by 
chance is less than 2\% so we are led to consider the possibility
that the medium may not be stationary in the LSR,
which we discuss later in Section 4.
We note that the uncertainty in screen location 
due to errors in the pulsar proper motion are
larger than those due to errors in the measured time scale.
Thus the two possible screen distances ($213\pm9$ pc and 178$\pm$12 pc)
are in fact jointly uncertain by about $\pm28$ pc.
However, since the pulsar velocity must be the same at each epoch
there is no single distance that matches both observations. 
It should also  be noted that, in any case, it is very unlikely that the
1998 observations could have come from the closer location,  
because that would require a very thin, very 
intense scattering screen. 

We also fit the simple Kolmogorov model with no inner scale
to the estimated power spectra. 
The 1998 data were best fit with a nonzero inner scale (as 
described in section 3.3) but this did not affect the estimate 
of the time scale.
The results, $\tau_{\rm w} = 1.84\pm0.22$ hr for 1996 and
$\tau_{\rm w} = 1.84\pm0.15$ hr for 1998, are not quite 
consistent with the time scales derived from the structure 
functions. The confidence limits for the two methods
just touch, suggesting that there may be a real bias.
This could occur because the power spectra and the 
structure functions emphasize different aspects of the data. 
The structure function fit is dominated by scales near $\tau_w$
whereas the power spectrum fit is dominated by the higher 
frequencies. The effect is that estimates from the power spectrum 
would move the location of the screen about 15 pc closer to the pulsar.
We use the time scales from the structure function in further analysis.

The plots of Figures \ref{fig:tweak1} and \ref{fig:tweak6}
are based on models in which the screen is at rest in the LSR.
In the Discussion (section 4) we consider models in which the
screen is moving with respect to the LSR, and compare these with
other astronomical information on motion in the local ISM. 
We now consider models with a uniform distribution of scattering.

\subsection{Extended Scattering Model}

When the scattering medium extends from the source to the observer,
the theoretical structure function is the sum of many screen
contributions each with an effective velocity $V_{\rm eff}(z_o)$
as in Appendix A.  One can visualize this by looking at
Figure \ref{fig:tweak1}; we add screen structure functions
weighted by the curve in the top panel, with each structure function
scaled in time by the heavy curves in the bottom panel.
We computed theoretical structure function for the 1996 and 1998 
observing times assuming that the scattering plasma is uniformly 
distributed along the line of sight and is stationary in the LSR.
As noted above
we reduced the time lags by a factor of 1.20 in each model,
to correct for the effect of incipient strong scattering.  
Then with the velocities known, the total variances $m_{\rm w}^2$ at
each epoch are the only free parameters in the models.
The corresponding least squares fits are plotted as dashed lines in 
Figures \ref{fig:lgstrfn}a and b.  These models have a more
rounded transition to saturation than for the screen geometry,
and they fit the observed time scales
reasonably well at half the saturation level. 
The $\chi^2$ values for these fits are 39 and 30 for
1996 and 1998 with 28 and 34 degrees of freedom, respectively. 
This shows that the extended medium model fits the data as 
well as the best fit thin screen and is intuitively more appealing 
as it has only one free parameter. 
We note that incipient strong scattering causes a slight change in shape
of the structure function as well as in  time scale; hence we cannot
use the differences in $\chi^2$ to discriminate between the models.

The LSR is based on the motion of nearby stars and it may not be the
best reference frame for the scattering plasma.  Though
differential Galactic rotation can be neglected over the 310 pc to the
pulsar, we have to consider what is known about motion of the nearby
interstellar gas.  Current views of the ``local bubble'' are described
in the proceedings of IAU colloquium 166 (Breitschwerdt et al 1998).
This ionized region within 100-200pc of the Sun must largely determine
the radio scattering for our pulsar observations.  However, the basic
idea of a hot ionized low density cavity has been refined by various
recent measurements.  Lallement (1998) discusses measurements of local
interstellar ``cloudlets'' within a few pc of the Sun moving at 5-10
km~s$^{-1}$ with respect to the LSR.  G\'{e}nova et al. (1998) describe the
kinematics of gas within the bubble using various tracers.  The
velocity of particular clouds are reported as much as 20 km~s$^{-1}$ relative
to the LSR; though it is not clear how these are related to the motion
of the radio scattering plasma which may extend along much of the line
of sight.


\begin{figure}[bth]
\psfig{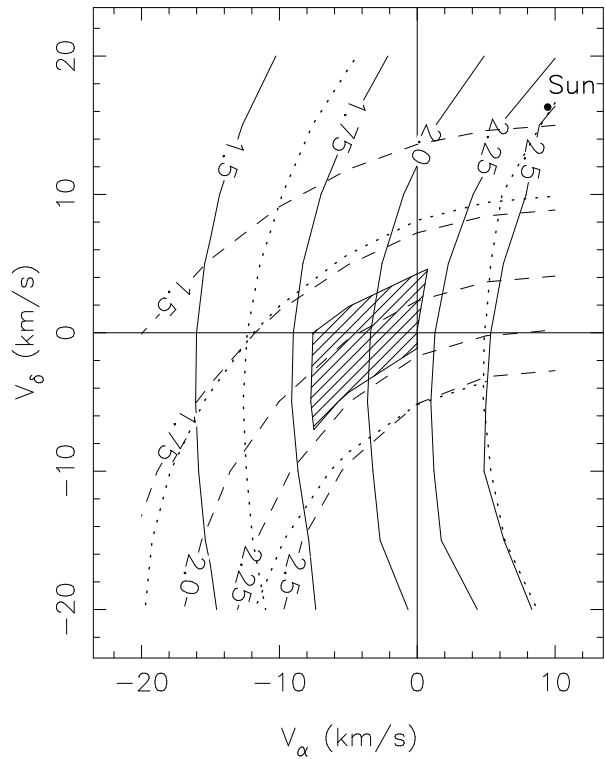}
\caption{ Scintillation time scale for observation in April 1996
(solid lines) and in October 1998 (dashed lines), 
calculated for a uniform scattering medium 
with a uniform velocity ($V_{\alpha}, V_{\delta}$) relative to the LSR. 
The time scales have been reduced by the factor 1.20
to account for the incipient strong scattering,
as described in the text. 
The hashed region is where the computed time scales agree 
with the observations within $\pm \sigma$ in both years. }
\label{fig:tauvel}
\end{figure}

Since these studies show a range of possible velocities in the local
ISM, we considered a model in which the scattering medium is
uniformly distributed between the Earth and the pulsar, but is
moving with respect to the LSR at constant velocity
($V_{\alpha}, V_{\delta}$). We performed an exhaustive search
in velocity space for a single medium velocity that simultaneously 
matches the time scales 1996 and 1998 data. 
Here the measured time scales were obtained from the thin
screen models fitted to the structure functions, since these models
provided a good template for the observations.

We present the results in Figure \ref{fig:tauvel}
as contours of the predicted time scales (again reduced by the factor 1.2)
in ($V_{\alpha}, V_{\delta}$) for the two observing 
periods.  The hashed region shows
where the calculated time scales lie within $\pm \sigma$ of the measurements.
Its extreme boundaries are $-8 \leq V_{\alpha} \leq 1$ km~s$^{-1}$ and
$-7 \leq V_{\delta} \leq 4$ km~s$^{-1}$ and just include the LSR itself.
The worst case errors due to uncertainties in the pulsar 
proper motion are shown by the dotted contours, which increases
the velocity range to $-12 \leq V_{\alpha} \leq 5$ km~s$^{-1}$ and
$-16 \leq V_{\delta} \leq 7$ km~s$^{-1}$. Since these are worst case
limits the associated confidence level is 90\%.
It would clearly be interesting 
to tighten the proper motion estimate.
We emphasize that this velocity only estimates the plasma
motion in a plane normal to the direction toward the pulsar
(at $l=140^{\circ}$, $b=31^{\circ}$) - plasma velocity along this
direction is not constrained.  
Upon reflection, it seems remarkable that 
the simplest uniform Kolmogorov model (with only one parameter -
the integrated strength of turbulence) provides a satisfactory
match to both observations.  We also note that the (projected) 
solar velocity in this frame lies nearly 20 km~s$^{-1}$ away and is
clearly excluded. 

To summarize, in the absence of other information about
the nature of the scattering medium,  we have two models that
explain the observations. A uniform scattering medium which moves 
at nearly the LSR velocity and a local scattering region at a 
distance of about 195 pc which is stationary in the LSR.  
It is also evident that localized
scattering regions at other distances would explain the
observations if they were moving at the right velocity.
We examine the tradeoff between distance and velocity in
the context of other astronomical information about the local
scattering environment in section 4.

\subsection{Spectra with an Inner Scale}

The possibility that the density spectrum is Kolmogorov but
includes an ``inner scale'' at which damping becomes
important, has been considered by several investigators
(e.g. Coles et al. 1987, Romani et al. 1986, Spangler and Gwinn, 1990).
The inner scale produces a rapid cutoff in the density spectrum
which can be modelled with a gaussian multiplier of the form
exp$[-\kappa^2 s_{i}^2/4]$. The intensity spectrum is directly
related to the density spectrum and shows the same multiplier.
However, the structure function shows the effect much less
dramatically - it changes exponent from
$D(s) \propto s^2$ to $D(s) \propto s^{5/3}$ above $s = s_i$ if
$s_i \ll r_f$.
We also note that both the screen and extended
medium curves in Figure \ref{fig:lgstrfn} approach the asymptotic
5/3 slope quite slowly; even the relatively
steep thin screen model has only reached a slope of 1.44 at
one tenth of the saturation level. Thus one
should be quite cautious about interpreting the
logarithmic slope of the structure functions
at small lags as estimates of $\alpha$.

We used fits to the power spectra, rather than the structure
function, to bound the inner scale. We tested both the thin
screen and uniform medium models described earlier.
The best fit theoretical spectra for thin screen and uniform 
medium models are shown in Figure~\ref{fig:pspec} 
as solid and dashed lines respectively. 
In 1996 all models with $s_i < 6 \times 10^6$ m fit equally well, 
whereas in 1998 the fits at $s_i = 2 \times 10^7$ m are 
significantly better. 
Although it is tempting to believe that an inner scale was 
detected in 1998, the detection depends on one or two points
at the highest frequencies in the observed spectrum.
These points are sensitive to the correction for system noise
and intrinsic variation. We have assumed that
both these corrections are white. A small amount of narrow
band noise could significantly bias the highest frequency
spectral estimates, increasing or decreasing them, depending
on whether the noise is above or below the data range.
Thus we have confidence only that the data indicate that 
$s_i \simless 2 \times 10^7$ m.

The only astrophysical plasma with which we can make a
comparison is the solar wind. There an inner scale has been
well determined (Coles and Harmon 1989), and appears to be 
very close to the ion inertial scale, i.e.\ the Alfv\'{e}n 
speed divided by the ion cyclotron frequency. Thus 
$s_i = 684\; (N_e$ cm$^{-3})^{-1/2}$ km depends only on the
density. At a typical density of $N_e = 0.03$ cm$^{-3}$ we 
have $s_i = 4 \times 10^6$ m.   
Spangler and Gwinn (1990) made a similar suggestion for the 
dissipation scale in the interstellar plasma, and provided
some evidence for inner scales $\simless 10^6$ m
for very heavily scattered lines of sight. 
It seems likely that the scattering
takes place in regions of somewhat higher density than the
mean interstellar medium, so it is likely that the inner scale
is less than $4 \times 10^6$ m. This is consistent with our
observations but the bound is not very tight. The dynamic range 
of the observations would have to be increased by about an order 
of magnitude to provide a more interesting bound.
An important consequence of the good spectral fits is to provide
new and independent evidence in suport of the Kolmogorov power
law for the spectrum of plasma density in the local ISM; this 
has become the canonical model, but is not often tested critically.

\subsection{Scintillation Index}

In the foregoing analysis we have concentrated on
interpretating the temporal variations characterized by
the normalized structure functions.  In weak scintillation
the time scale depends only on the relative distribution
of turbulence along the line of sight, and is independent of
its absolute level. We now estimate the absolute level
of turbulence from the measured scintillation index
and compare it with other observations of the same pulsar
and also with models for the ISM electron density.
The models describe the distribution of the
variance of density, parameterized by
$C_N^2$, the constant in equation
(\ref{eq:kolmo}) for the density spectrum.
But, in general, the observations can only estimate
the integral of $C_N^2$ along the line of sight,
termed the scattering measure $SM$.

Observations of ISS have shown extreme variability
in $SM$ between even rather closely aligned sight lines
Previous workers (e.g.\ Cordes, Weisberg and
Boriakoff, 1993 [CWB]; TC93) have modelled the distribution of
$C_N^2$ as a locally uniform background, with a stronger
and more clumpy component superimposed.
These models have been based primarily on pulsar
measurements of the diffractive decorrelation bandwidth
$\delta \nu_d$ in strong scattering.  Though this
parameter is easy to estimate over 1-2 hours, the repeated
327 MHz observations of Bhat et al.\ (1999a) show it
to be surprisingly variable. For many of the 20 pulsars
observed within 1 kpc, $\delta \nu_d$ varied by
a factor of 5 or more over days to weeks.
This variation is almost
certainly caused by refractive scintillation,
which modulates the diffractive spectrum and is caused by
much larger scales in the ISM than those responsible for the
diffractive scintillation itself.
As noted by Cordes, Pidwerbetsky and Lovelace
(1986), and by Gupta et al.\ (1994),
the refractive modulations also tend to bias the apparent
$\delta \nu_d$ downwards, suggesting that
the larger values of $\delta \nu_d$ more accurately
estimate the diffracting irregularities.   With this preamble, we now
compare $SM$ derived from our scintillation index
measurements with estimates based on $\delta \nu_d$
measured for the same pulsar and also with $SM$ from
the TC93 and Bhat et al.\ (1998) models.

Our observations of the scintillation index
give $SM \sim 3 \times 10^{-6}$m$^{-20/3}$kpc,
depending slightly on whether we
assume a screen or uniform scattering
medium (see Table 1). Also listed are
$SM$ estimated from two $\delta \nu_d$ observations
(at 151 MHz by Rickett, 1970 and at 360 MHz
by Cordes, 1986). We assume a uniform scattering medium
in estimating $SM$ from these measurements, and
we noticed that Cordes (1986)
used a formula higher by a factor 3.2 than
that in TC93.  As discussed in Appendix C,
we use the formula in TC93, since it agrees
with the independent analysis of Lambert and Rickett (1999).
Even with this correction, the 151 MHz observation gives
an $SM$ 3.6 times higher than ours and the 360 MHz one is 13 times
higher. The two measurements of $\delta \nu_d$ differ
by a factor 5 when scaled to a common frequency.
From the discussion above, we may assume that the
larger value of $\delta \nu_d$,
i.e.\ the smaller value of $SM$,
better represents the underlying density
spectrum responsible for small scale ISS;
but even then we have a factor of 3.5 discrepancy in $SM$.
Both methods scale in the same way
with screen distance and with pulsar distance,
and so they cannot be reconciled by
adjusting the distances in either the screen
or uniform scattering geometries.
In agreement with Bhat et al.\ (1999b), we conclude that
we need a better understanding of the variability in
$\delta \nu_d$ observations; numerical simulations
will be necessary, since the existing theory assumes asymptotically strong
scattering which is not strictly applicable on the shorter
paths to nearby pulsars.

\begin{table*}[htb]
\tablecaption{Scattering Measure Estimates towards PSR B0809+74}
\begin{tabular}{ccccc}
\tableline
Observation & Frequency & Reference & Model & $SM$  \\
 & (GHz) &  &  & (m$^{-20/3}$kpc) \\
\tableline
$m_{\rm w}=0.8$ & 0.93  & this work & screen at 195 pc & $2.3 \times 10^{-6}$\\
$m_{\rm w}=0.8$ & 0.93 & this work & extended med. & $3.0 \times 10^{-6}$\\
$\delta \nu_d = 0.95$ MHz \tablenotemark{a} & 0.36 & Cordes(1986) &
extended med. & $4.1 \times 10^{-5}\ast $\\
$\delta \nu_d = 0.1$ MHz \tablenotemark{b} & 0.15  & Rickett(1970) &
extended med. & $1.1 \times 10^{-5}\ast $\\
Theory &  & TC93 & extended med.
& $7.1 \times 10^{-5}$\\
Theory &  & Bhat et al.\ (1998) & shell of bubble
& $5.5 \times 10^{-5}\ast $\\
\tableline
\end{tabular}
\tablecomments{* $SM$ obtained using the formulae in Appendix C (and in TC93)}
\tablenotetext{a}{average of observations at 0.36 and
0.41 GHz scaled to .36 GHz}
\tablenotetext{b}{ Value obtained from measurement
of bandwidth that reduces scintillation index to 0.5, divided
by 10 (see CWB)}
\end{table*}

We also calculated $SM$ from the models of TC93 and Bhat et al.\ (1998).
The former is 23 times and the latter is 18 times our estimate.
TC93 model the plasma as uniform turbulence in the Galactic disk
of 700 pc scale height in $C_N^2$.
The recent model proposed by Bhat et al.\ (1998) adds
detail near the Sun of a bubble of low turbulence
surrounded by enhanced turbulence in a shell at 50-150 pc.
They propose a range of ellipsoidal models for the shell, which
the line-of-sight to PSR B0809+74 ($l=140^{\circ}, b=31.6^{\circ}$)
crosses at $72 \pm 13$ pc from the Earth.
The model specifies a low $C_N^2$ inside the bubble,
a value of $SM$ for the shell and the same $C_N^2$
as TC93 outside the shell. For our 310 pc path their
shell contribution to $SM$ dominates with 70\%  and the outer
region provides 30\% and the inner region less than 1\%.
Their model is closer to being a ``screen'' than a uniform extended
scattering medium. Since we find 18 times lower
$SM$ than in their model, we conclude that the enhanced
turbulent shell is not continuous and must have a hole toward
pulsar B0809+74.
We note that of the 20 pulsars observed by Bhat et al., the
closest to PSR B0809+74 was PSR B0823+26 at the same
Galactic latitude but 50 degrees away in longitude.
Thus their observations are not in conflict
with our low $SM$ toward pulsar B0809+74.

The large $SM$ discrepancies with the scattering models
could be resolved by substantially reducing the pulsar
distance. If $L$ were reduced by a factor 5.5 it would
bring our $SM$ into line with that from TC93 and bubble models.
If that were the case, at 56 pc this would make PSR B0809+74 the closest
pulsar to the Earth and the mean electron density 0.1 cm$^{-3}$,
3 to 5 times higher than other estimates. For example,
PSR B0950+08 has a measured parallax distance of 130 pc
and a mean electron density 0.025 cm$^{-3}$.  Though
the distance  to B0809+74 may be uncertain by as much as 50\%,
it cannot be reduced by the factor 5.5, and
we conclude that the turbulence level on this line of sight
is much lower than described by either of the models.
From our estimate of $SM$
the effective average $C_N^2$ toward B0809+74
$\sim 10^{-5}{\rm m}^{-20/3}$.
This is comparable to that in the inner cavity of the bubble
model. It also agrees with that toward PSR B0950+08
($l=229^{\circ}$, $b=44^{\circ}$)
from the measurements of Phillips and Clegg (1992),
after applying the correction factor from Appendix C.
The low $C_N^2$ for both of these pulsars suggests a quite
large region of low turbulence in the hemisphere
away from the Galactic center. The direction to our pulsar looks
out from the local arm through an inter-arm region,
which we suggest may have a lower plasma density than that described by the 
TC93 model. The low density might also be related to the
local bubble, except that the X-ray observations of Snowden (1990)
indicate that the bubble terminates between 50 and 100 pc towards 
PSR B0809+74. 

\section{Discussion}

We have found that the observations agree with the predictions
for a Kolmogorov density spectrum in either
a uniformly extended scattering medium moving at 
close to the LSR velocity or in screen at rest in the LSR,
located in the range 170 -- 220 pc from the Earth.
We now consider whether these results are consistent
with other ideas about the local scattering environment.


\begin{figure}[bth]
\psfig{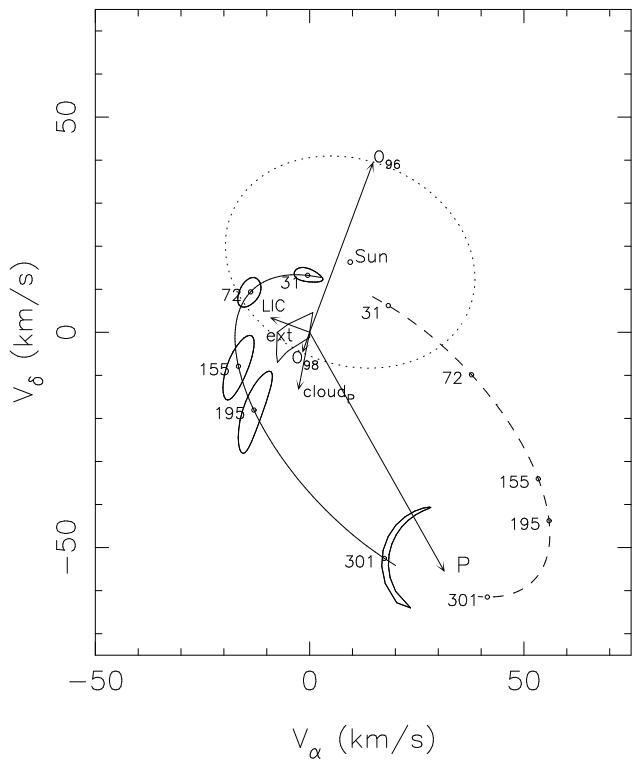}
\caption{Velocities relative to the LSR, projected on the
plane normal to the line of sight toward pulsar B0809+74.
The pulsar velocity is marked by P.
The Earth's velocity is shown for the observations in 
1996 and 1998, lying on the dotted ellipse of the 
Earth's annual track, centered on the solar velocity.
For each observation the line of sight cuts through the medium
at a velocity that lies on a line between the Earth and pulsar
velocities. ``ext'' marks 1 $\sigma$ region of allowed velocity for 
a uniform scattering medium that best matches the data.
Screen models give pairs of solutions for the screen
velocity, which follow partial ellipses. One solution 
is drawn dashed since its larger velocities make it less likely.
At 5 screen distances error contours 
are drawn for the allowed screen velocity.   The plot does not include
errors due to uncertainties in the proper motion.
Also shown are the projected velocities of two relevant 
interstellar clouds in the solar neighborhood as discussed 
in the text.}
\label{fig:velplot}
\end{figure}

In the local bubble model of Bhat et al.\ (1998),
enhanced scattering is expected at $72 \pm 13$ pc from the Earth.
This is marked on Figure \ref{fig:tweak1} by a rectangular bar
and is clearly inconsistent with the screen solution near 195 pc.
If the scattering region were located at 72 pc and
co-moving with the LSR, the predicted
time scale would be about 30\% longer than observed in 1996
and twice as long as observed in 1998.
Consequently, we also examined screen models with a fixed
position and a free velocity $\bm{V}_{\rm scr}$ relative to the LSR.

We have measured weak scintillation time scales
($\tau_{\rm w96}$ and $\tau_{\rm w98}$) for two epochs.
If $x=z_o/L$ denotes the fractional distance to the screen,
then the results of Appendix A give
$\tau_{\rm w} = 0.98 \sqrt{\lambda Lx(1-x)/2\pi}/
|(1-x)\bm{V}_o+x\bm{V}_p-\bm{V}_{\rm scr}|$.
With $\bm{V_o}$ appropriate for the two epochs there are
two equations, from which two possible solutions for
$\bm{V}_{\rm scr}$ can be found at a given $x$.
In Figure \ref{fig:velplot} these two velocity solutions are plotted
as partial ellipses (solid and dashed lines), annotated with 
sample distances. The LSR velocity of 
a screen necessary to match the two
observed time scales is thus given by a vector from the origin to
the appropriate point on either curve. 
The lower velocity solutions are shown with sample
error ellipses corresponding to the
uncertainties in the time scale estimates.
There are also errors due to uncertainty in the
pulsar proper motion, given moving the point P
by $\pm10$ km~s$^{-1}$ in either coordinate.
These map in a non-linear fashion into 
an error in $\bm{V}_{\rm scr}$, which is
$\sim 10$ km~s$^{-1}$ for screens near the pulsar, decreasing
for screens nearer the Earth. 

A screen at the proposed bubble distance of
72 pc needs a particular velocity
of $V_{\alpha}=-13.8 \pm 3, V_{\delta}=9.4 \pm 4$ km~s$^{-1}$ 
transverse to the pulsar line of sight.  While 
this is a reasonable interstellar velocity, the 
problem remains that our estimate of $SM$
is 18 times smaller than in the Bhat et al. model.
The solution from section 3.1 with a screen near 195 pc
requires the velocity $V_{\alpha}=-13 \pm 4, V_{\delta}=-18 \pm 10$ km~s$^{-1}$
indicated by the elongated error ellipse, 
rather than being at rest in the LSR. 
We have also marked as a parallelogram the (modest) velocity 
needed for scattering uniformly extended along the entire 
line of sight.  The pulsar B0809+74 happens to 
lie at high ecliptic latitude
and so the projected earth velocity (shown by the dotted ellipse
in Figure \ref{fig:velplot}) samples a large region of
velocity space.  It is likely that we can further constrain the
scattering region by a series of weak ISS measurements
at 2-3 month intervals over the year.

Figure \ref{fig:velplot} is also useful for displaying
other velocity measurements of the local ISM, two of
which we have projected onto our line of sight
and plotted as vectors.  The arrow marked LIC
is the projected LSR velocity of the local interstellar cloud
observed to be entering the heliosphere from 
its backscattered Lyman alpha (see Lallement, 1998).  
This cannot be related to the bubble solution with a 
comparable velocity, since the bubble model has
no scattering nearer than about 70 pc. 
G\'{e}nova et al.\ (1998)
report measurements of Na I absorption
lines over a wide range of galactic longitudes.
They identify an interstellar cloud with velocity
of 13.8 km~s$^{-1}$ (toward $l=225^{\circ}, b=5.4^{\circ}$)
covering a large solid angle including  PSR B0809+74.
The arrow marked ``Cloud P'' is the projected velocity of this
interstellar cloud, which is comparable to that of our
sample screen solution at 195 pc.  
G\'{e}nova suggests (private communication 1999)
that it may be a large warm diffuse cloud that
extends 70 to 120 pc. Of course the other question in 
such a comparison is that
the optical absorbing cloud may not be physically the same as
the ionized region of enhanced turbulence.
We also note that the likely interstellar plasma
velocities are relatively small.
The Alfv\'{e}n speed in the warm ionized medium, which is
assumed to be responsible for the scattering, is
thought to be about 10 km~s$^{-1}$, similar to that of cloud motions.

Britton et al. (1998) have compared the angular scattering of
pulsars with their temporal broadening and concluded that the
scattering is often close to the pulsar. They suggest that it 
may be due to a nebula associated with each pulsar. This
scenario cannot explain our observations of B0809+74, primarily 
because it is thought to be relatively old, of order
T = 10$^8$ yrs. If the scattering takes place at a distance R
from the pulsar then the velocity of the scattering nebula $V_{neb}$ 
with respect to the pulsar must be less than R/T. 
This is a very tight bound and such low velocities would
produce much longer time scales than we observe.
For example if R = 10 pc, then $V_{neb} < $0.1 km~s$^{-1}$. However
from Figure 7 we can see that a screen at a distance $L-R$ = 301 pc
from the Earth must be moving at 12 to 14 km~s$^{-1}$ with
respect to the pulsar to explain our measured time scales. 
For scattering near the pulsar the time scale can be approximated by
$\tau_{\rm w}' = 0.82 T / \sqrt{k R}$, which is 250 hrs at R = 10 pc.
Evidently a model of scattering near the pulsar would only be 
tenable only if the age of B0809+74 were less than 
$0.6 \times 10^6$ yrs.

Our results can also be used to predict the
interstellar scintillation for extragalactic sources
B0716\-+714 and B0836\-+710, which are within $5^\circ$ of B0809\-+74.
Evidently, the predictions depend on whether or not we extrapolate
the low level of scattering that we measure over 0.3 kpc
out over the entire 1 kpc pathlength of the TC93 model.

For example B0716+714 is known to show rapid optical and radio
variations (Wagner et al. 1996). The optical variations cannot 
be due to ISS and so imply a very small optical size; if this same 
size applied at cm wavelengths then it must also scintillate
as an effective point source in the interstellar medium. 
Extrapolating our low $C_N^2$ over the entire
pathlength of the TC93 model, 
we predict $m_w \sim 0.4$ , and spatial scale 
$\sim 7 \times 10^8$m at 5 GHz. 
Because the transverse LSR velocity of the earth varies considerably
in this direction, the associated time scale $\tau_w$ would 
vary from 50 hrs in October to 5 hrs in April.
With the higher $C_N^2$ level in the TC93 model
the point source $m_w \sim 2.1$, indicating strong 
scintillation, which would split into diffractive
and refractive components; however, their spatial and temporal
scales would not be very different from the estimates above.
A more realistic source model would include a maximum brightness
(minimum angular size), which would reduce the rms flux variation
and increase the time scales; nevertheless the strong annual
variability in time scale would remain, if the
scattering plasma does indeed share the motion of the LSR
out to 1 kpc.

The source B0836+710 is known to have a very compact component
from 20 GHz VLBI observations (Otterbein et al., 1998). 
If the diameter observed at 20 GHz applied at 5 GHz, it should
also show interstellar scintillations, which are not observed
(Quirrenbach, et al., 1992). However, at 5 GHz
the compact component appears to be self-absorbed, so predicting 
the expected scintillation index versus frequency will require a
full model for the source structure versus frequency.

\section{Conclusions}

We have developed a method for observing multiple epochs
of weak scintillation of nearby pulsars whose proper motion
has been measured. Our observations for pulsar
B0809+74 can be explained by a uniform distribution of 
plasma ``turbulence'' which is stationary in the LSR.
The density spectrum follows Kolmogorov law for
scales between $2\times 10^7$m and $10^9$m, i.e.\
the turbulence may have an inner scale
$\simless 2 \times 10^7$ m. This bound is
consistent with the hypothesis that the inner
scale is the ion inertial scale, which would be about
$4 \times 10^6$ m for an electron density of 0.03 cm$^{-3}$.
The velocity of the scattering medium that best fits the
observations is in the range $-12 \leq V_{\alpha} \leq 5$ km~s$^{-1}$ and
$-16 \leq V_{\delta} \leq 7$ km~s$^{-1}$. The error bounds, which are 
dominated by error in the pulsar proper motion, include the LSR.

The scintillation is also consistent with scattering from a
localized region which is not stationary in the LSR, but this 
appears to be less likely for two reasons. 
First the scattering is considerably weaker than 
predicted by models of the distribution of turbulence in the local
neighbourhood, so it is unlikely that the line of sight to
B0809+74 is dominated by a local intensely scattering region.
Second, if the scattering is dominated by a local region, it
must be moving at a substantial velocity with respect to the
LSR as shown in Figure~\ref{fig:velplot}.
While this is possible, it appears to be somewhat less likely.
The scintillation is not consistent with scattering in a
``nebula'' associated with the pulsar because the time scale
of the scintillation is too short by several orders of
magnitude.

\acknowledgements

The authors would like to acknowledge the help of Michael
Kramer who pointed out that pulsar B0809+74 has a substantial 
polarization. This lead to a reanalysis that
significant changed the conclusions and improved this
paper.  We thank the director and Staff of EISCAT for
the use of the Sodankyl\"{a} Telescope and we thank the NSF for partial
support under grant AST 94-14144.

\appendix

\section{Weak Scintillation Theory}
\label{sec:weakscint}

We give here the theory of weak interstellar scintillation of a
point source (pulsar) moving with respect to a scattering
medium observed by a moving observer. Since the scattering is
weak we can calculate the contribution of a single thin screen
and use the Born approximation to add the effect of many thin
screens linearly.

Consider a thin phase screen between a point source P
and an observer O. The screen is at distance $z_p$ from P
and $z_o$ from O with $z_o + z_p = L$.
The screen introduces a phase change $\phi({\bm s_1})$
at transverse coordinate ${\bm s_1}$.  Then
the Fresnel diffraction integral for the complex field at
an observer (0,0,$L$) can be written:
\bea
f(0,0,L) \simeq  \frac{jk}{2 \pi z_e} \exp [-jkL]
\int d^2{\bm s}_1 \nonumber \\
\exp \left [ j \phi (\bm{s}_1) -j \frac{k s_1^2}{2 z_e}
\right ]
\label{eq:f00L}
\eea
where $z_e$ is defined by $1/z_e = 1/z_o + 1/z_p$
and $k$ is the radio frequency propagation constant.
The factor $jk/2 \pi z_e$
ensures that the field has unit average intensity ($<ff^*>$) at $z=L$.

If now we consider the effect of moving the source to a
transverse coordinate ${\bm s_p}$ and the observer
to ${\bm s_o}$, which are small compared to $z_p$ and $z_o$,
the field can be expressed as:
\bea
f({\bm s_p},{\bm s_o},L) \simeq  \frac{jk}{2 \pi z_e} 
\exp [-jkL +j \Phi({\bm s_p},{\bm s_o}) ] \times \nonumber \\
\int d^2{\bm s_1} \;  \exp \left [ j \phi ({\bm s_1}) -j
\frac{k |{\bm s_1}-{\bm s_{\rm eff}}|^2}{2 z_e} \right ]
\label{eq:fssL}
\eea
Here
\bea
\Phi({\bm s_p},{\bm s_o}) = k |{\bm s_p}-{\bm s_o}|^2 / 2L
\nonumber \\
{\bm s_{\rm eff}} = {\bm s_p} z_o/L + {\bm s_o} z_p/L ,
\label{eq:seff}
\eea
where $s_{\rm eff}$ is the transverse coordinate where a straight line from
P to O
intersects the screen.

From eq (\ref{eq:fssL}) we can find the intensity
$I({\bm s_p},\-{\bm s_o},L) = ff^*$ as a double integral.
We need the intensity covariance
\bea
R_I^s({\bm \sigma_p},{\bm \sigma_o},L)  =
\left< \; I({\bm s_p},{\bm s_o},L)  \right. \times \nonumber \\
\left. I({\bm s_p}+{\bm \sigma_p},{\bm s_o}+{\bm \sigma_o},L)
\; \right> 
\label{eq: RissL}
\eea
Substituting the double integral form of $I$ into the above equation
generates the ensemble average of a quadruple integral over
transverse screen variables
(${\bm s_1},{\bm s_2},{\bm s_3},{\bm s_4}$). After transforming to
variables which are the various sums and differences of these
and performing two integrations we obtain the spatial
covariance
\bea
R_I^s({\bm \sigma_p},{\bm \sigma_o},L)  =
\frac{k^2}{(2 \pi z_e)^2}  \int d^2{\bm r} \;\int d^2{\bm q} \nonumber \\
\exp \left [- 0.5 D_4({\bm r},{\bm q}) -
\frac{jk}{z_e}({\bm r}\cdot{\bm q} +
{\bm \sigma}_{\rm eff}\cdot{\bm q}) \right ] 
\label{eq:RissLint.gen}
\eea
where  ${\bm r} = ({\bm s_1}+{\bm s_2}-{\bm s_3}-{\bm s_4})/2$
and  ${\bm q} = ({\bm s_1}-{\bm s_2}+{\bm s_3}-{\bm s_4})/2$
and $D_4({\bm r},{\bm q}) = D({\bm r}) + D({\bm r}) -
0.5 D({\bm r}+{\bm q}) -0.5 D({\bm r}-{\bm q})$
and ${\bm \sigma}_{\rm eff}$ is defined in an equation analogous to 
\ref{eq:seff}. Here the function $D({\bm r})$
characterizes the phase screen by the structure function
of its phase at offset ${\bm r}$.  Standard analysis of a plane wave
normally incident on a phase screen gives the
correlation function for intensity
at spatial offset ${\bm s}$ at a distance z from the screen
(e.g. equation (5.3) of Tatarskii and Zavorotnyi, 1980).
Equation (\ref{eq:RissLint.gen}) is related to this simpler result by
replacing ${\bm s}$ by ${\bm \sigma}_{\rm eff}$ and $z$ by $z_e$.
A general treatment for waves scattered by such a screen
requires a separate
analysis for weak and strong scintillation. Under weak scintillation
the important regions in ${\bm r}$ and ${\bm q}$ that dominate
the integrations are where $D_4$ is small compared to one,
and $\exp[-0.5 D_4]$ can then be expanded to first order and
integrated over ${\bm r}$.  The result
is often called the Born approximation [e.g equation (4.11)
of Prokhorov et al. (1975)].

We are concerned with temporal variations caused as the observer moves
at velocity ${\bm V_o}$ and the pulsar moves at ${\bm V_p}$,
both velocities with respect to the screen. The observer measures
a temporal correlation function at time offset
$\tau$, which is given by:
\be
R_I^t(\tau) = R_I^s({\bm \sigma_p } = {\bm V_p}\tau ,
{\bm \sigma_o}={\bm V_o}\tau ,L)
\label{eq:Rittau }
\ee
The associated structure function is
\be
D_I^t(\tau) =  2 R_I^t(0) - 2 R_I^t(\tau)
\label{eq:Dittau}
\ee
\noindent
The weak scintillation Born approximation gives:
\bea
R_I^t(\tau) = \int d^2{\bm \kappa} \;
 P_{\phi}({\bm \kappa}) \; 4 \sin^2 (\kappa^2 z_e/2 k) 
\times \nonumber \\
\exp[-j {\bm \kappa}\cdot{\bm V_{\rm eff}}\tau ]
\label{eq:Rittau.screen}
\eea
Here $P_{\phi}({\bm \kappa})$ is the spectrum of the
phase introduced by the screen (ie for a normally incident plane wave)
at wavenumber ${\bm \kappa}$, which corresponds to $(k{\bm q}/z_e)$
in the notation used above, and
\be
\bm V_{\rm eff} = {\bm V_o} z_p/L + {\bm V_p} z_o/L
\label{eq:veff}
\ee
which is the effective velocity of the point where a straight
line from the pulsar to the observer intersects the screen.
Note also that the low-wavenumber approximation in strong scattering
is also given by this result with an extra exponential
factor inside the integral $\exp\-[-D({\bm \kappa} z_e/k)]$.
For a layer of thickness $\Delta$,
we replace $P_{\phi}({\bm \kappa})$ by
$P_{\phi}'({\bm \kappa})\Delta$, where
\be
P_{\phi}'({\bm \kappa}) = 2 \pi r_e^2 \lambda^2
P_{N_e}(\kappa_x,\kappa_y,\kappa_z=0,z_o)
\label{eq:P_phi'}
\ee
where $\lambda = 2 \pi/k$ and
$P_{N_e}(\kappa_x,\kappa_y,\kappa_z,z_o)$ is the three-dimensional
power spectrum of the electron density in the layer
at a distance $z_o$ from the observer.
If $P_{\phi}'({\bm \kappa})$ is isotropic we obtain
\bea
D_I^t(\tau) = 32 \pi^2 r_e^2 \lambda^2 \Delta
\int_{0}^{\infty}  \kappa d \kappa \;
P_{N_e}(\kappa , z_o) \times \nonumber \\
\sin^2 (\kappa^2 z_e/2 k)
\left [ 1 - J_0(\kappa |V_{\rm eff}| \tau ) \right ] .
\label{eq:Dittau.J0.screen}
\eea
This equation is used to compute models for comparison
with the observations in section 3.  With a simple power law
density spectrum,
\begin{eqnarray}
P_{N_e}(\kappa , z_o) = C_N^2(z_o) \kappa^{-\beta}  \label{eq:kolmo}\\
\beta = 11/3 , \; \; \; \alpha = \beta -2, \nonumber
\end{eqnarray}
the observed temporal structure function can be expressed as
\be
D_I^t(\tau,z_o) = 2 m_{\rm w}^2 \; d_{\rm scr}(\rho )
\label{eq:Ddscreen}
\ee
where
\bea
d_{\rm scr}(\rho) & = & g_{\alpha} \; \int_{0}^{\infty} y^{-1-\alpha}
\sin^2(y^2/8) \times \nonumber \\
& & [1-J_0(y \rho /2)] \; dy  \\
\rho & = &  \tau V_{\rm eff}\sqrt{k/z_e(z_o)} \nonumber \\
g_{\alpha} & = & \frac{\alpha 2^{1+\alpha/2} \Gamma(1/2+\alpha/4) }
{\sqrt{\pi} \Gamma(1-\alpha/4) }  \nonumber
\label{eq:dscr}
\end{eqnarray}
With these definitions $d(\infty) = 1$, and
for the Kolmogorov spectrum ($\alpha=5/3$),
we find $d_{\rm scr}(0.99) = 0.5$ and so the scintillation time scale
is $\tau_{\rm w} = 0.98 \sqrt{z_e/k}/V_{\rm eff}$, and
$g_{\alpha} = 7.346$.
We also give expressions for the scintillation index in terms of the screen
coherence scale $s_{\rm scr}$ as
\begin{eqnarray}
m_{\rm w}^2  =  \frac{\pi \alpha/4} 
{\Gamma(1-\alpha/2) \sin(\pi \alpha/4)} \times \nonumber \\
\left [ \frac{ 2 \sqrt{z_e(z_o)/k}} {s_{\rm scr}} \right ]^{\alpha} \\
s_{\rm scr}^{-\alpha}  =  8 \pi^2 r_e^2 \lambda^2 
C_N^2(z_o) \Delta \times \nonumber \\
\frac{\Gamma(1-\alpha/2)} {\alpha 2^{\alpha} \Gamma(1+\alpha/2)}
\nonumber
\label{eq:s_scr}
\eea
Here the coherence scale for the screen $s_{\rm scr}$ is
defined as the lateral spatial offset over which
there is an rms difference of 1 radian in the screen phase.

For an extended medium in weak scintillation we find
$D_I^t(\tau)$  by simply integrating along the line of sight,
where the screen thickness $\Delta$ becomes $dz_o$.
We proceed by assuming the medium to be ``frozen'' and refer
${\bm V_o}$ and ${\bm V_p}$ to the rest frame for the medium.
In completing the integral, we assume that $C_N^2$ is independent of $z_o$
but include the $z_o$ dependence of the other variables:
\begin{eqnarray}
& & D_I(\tau,V_p,r,\theta_p) = \int_0^L  D_I^t (\tau, z_o) dz_o/ \Delta 
\nonumber \\
& & z_e(z_o) =  z_o (L-z_o)/L  \\
& & V_{\rm eff}(z_o)  =  V_p \; [ z_o^2  +
2 z_o(L-z_o) r \cos \theta_p  + \nonumber \\
& & (L-z_o)^2 r^2 ]^{1/2} /L \nonumber
\label{eq:D_ext}
\end{eqnarray}

The net effect is that each screen of thickness $dz_o$ contributes
with a different time scale to the total signal. This rounds the abrupt
transition to saturation that is found in the thin screen.
The variation of $V_{\rm eff}$ is determined by
$r = V_o/V_p$ and $\theta_p$ the angle
between $\bm{V_p}$ and $\bm{V_o}$. Thus the detailed shape
of the resulting structure function depends on the parameters
$r$ and $\theta_p$.  At one extreme equal and parallel velocities
($r=1, \theta_p = 0$), $V_{\rm eff}$ is independent of distance
and the structure function is just a stretched version of
$d_{\rm scr}$.  In the case of anti-parallel velocities
$V_{\rm eff}$ covers a wide range since
there will be a distance at which it goes to zero and the
corresponding time scale $\tau_w \rightarrow \infty$.
In our application we know the pulsar and observer velocities
with respect to the Sun.  Our basic model assumes
that the medium (either screen or uniform extended medium)
moves with the local standard of rest.
But the theory can equally be applied to models with the medium
in uniform motion, if $\bm{V_p}$ and $\bm{V_o}$ represent the
velocities with respect to the medium.

\section{Error in Structure Function Estimates}
\label{sec:strerror}

We need an expression for the errors expected in estimates of
the structure
function, in terms of the time lag $\tau$ and duration of the
data sequence $T$ and the characteristic time constant for the process
$\tau_o$.  We present here an analytical solution and a simulation.
The estimator used can be written in continuous (symmetrical) form as
\be
\hat{D}(\tau) = \int_{-T'/2}^{T'/2}  \Delta X(t,\tau )^2 dt/T'
\label{eq:dhat}
\ee
where  $\Delta X(t,\tau ) = X(t-\tau /2) - X(t+\tau /2)$ and
$T' = T - \tau$.  This estimator is unbiased as:
\be
\langle \hat{D}(\tau) \rangle  = \langle \Delta X(t,\tau)^2
\rangle  \; = {D}(\tau),
\label{eq:dhat_ave}
\ee
which is independent of $t$ or $T$.
The covariance in the estimators at time lags $\tau_1$ and
$\tau_2$ is then
\bea
{\rm Cov}[ \hat{D}(\tau_1),\hat{D}(\tau_2) ] = \nonumber \\
\langle \hat{D}(\tau_1) \hat{D}(\tau_2) \rangle -
\langle \hat{D}(\tau_1) \rangle  \langle \hat{D}(\tau_2) \rangle  .
\label{eq:covdef}
\eea

Substituting \ref{eq:dhat}
and taking the ensemble average inside the integrals
this yields
\bea
{\rm Cov}[ \hat{D}(\tau_1), \hat{D}(\tau_2) ] = 
\int_{-T_1'/2}^{T_1'/2}  \int_{-T_2'/2}^{T_2'/2}  \nonumber \\
\langle  \Delta X(t_1,\tau_1)^2  \Delta X(t_2,\tau_2)^2  \rangle \nonumber \\
dt_1 dt_2/T_1'T_2'  -  D(\tau_1) D(\tau_2) ,  
\label{eq:covintegral}
\eea
where $T_1'=T-|\tau_1|$ and a similar expression for $T_2'$.
We follow the method used by Jenkins and Watts (1969, pp 412-3, - JW)
and assume that $\Delta X$ is a Gaussian random variable with
zero mean.  Using the well-known average of the product of
four such variables, we obtain:
\bea
{\rm Cov}[ \hat{D}(\tau_1), \hat{D}(\tau_2) ] =
\int_{-T_1'/2}^{T_1'/2}  \int_{-T_2'/2}^{T_2'/2} \nonumber \\
2 \langle \Delta X(t_1,\tau_1)  
\Delta X(t_2,\tau_2)   \rangle^2  dt_1 dt_2 / T_1'T_2'  .
\eea
Following (JW), we change to sum and difference variables of integration.
The region of integration becomes a parallelogram,
over which the sum variable is directly integrable
to give
\begin{eqnarray}
{\rm Cov}[ \hat{D}(\tau_1), \hat{D}(\tau_2) ] =   \nonumber \\
\int_{0}^{T''}  4 \Gamma(r,u,v)^2  (T''-|r|) dr / (T''^2-v^2) \nonumber \\
-  \int_{0}^{v} 4 \Gamma(r,u,v)^2 (1-|r|/v) dr/ (T''-v) ,
\label{eq:cov-uv}
\end{eqnarray}
where $u=(|\tau_1|+|\tau_2|)/2$, $v=||\tau_2|-|\tau_1||/2$  and
$T'' = (T_1'+T_2')/2 = T-u$, and
\bea
\Gamma(r,u,v) = 0.5 [ D(r-u) + D(r+u) - \nonumber \\
D(r-v) - D(r+v) ]   .
\label{eq:Gammdef}
\eea

\subsubsection*{Variance in structure function with power law behavior}

The interesting special case of the variance in $\hat{D}(\tau )$
corresponds to $v=0$ $u=\tau$ and $T'' = T'$,
for which the second integral in equation (\ref{eq:cov-uv}) is zero.
We proceed by considering the various asymptotic forms.
For small lags, $D$ increases (monotonically) with lag
and saturates at $D_{\infty} = 2 \sigma_X^2$. A
characteristic time scale separates these regions and can be defined
as $D(\tau_o) = 0.5 D_{\infty}$.
There are limiting cases in which the observing sequence is
long enough to include many independent variations ($T \gg \tau_o$)
and the opposite extreme ($T \ll \tau_o$).

First, consider $T \gg \tau_o$. Then with $T > \tau \gg \tau_o$,
the triangular weight in eq (\ref{eq:cov-uv})
can be ignored and the upper limit of the integral can be extended
to infinity:
\be
{\rm Var}[\hat{D}(\tau)] \sim (4/T') \int_0^{\infty} \Gamma(r,\tau,v=0)^2 dr  .
\label{eq:vard}
\ee
This gives the expected behavior for rms errors in
$\hat{D}$ as proportional to $\sqrt{\tau_o/T'}$.
At large lags $\tau \simgreat \tau_o$, $D$ then approaches saturation and
\be
\Gamma(r,\tau,v=0) \sim   D_{\infty} - D(r)  = 2 R(r)
\ee
where $R(r)$ is the covariance function is related to $D(r)$ as indicated.
When put back into equation(\ref{eq:vard}), the result is 4 times
the variance in the conventional autocovariance estimator at zero lag
$\hat{R}(0)$, using the standard formula given
by JW equation (5.3.21) with lag equal zero.
We can then see that the two methods for estimating $\sigma_X^2$
have the same rms error.
Having discussed the errors for large lags, we now address small lags
$\tau \ll \tau_o \ll T $.  Then $\Gamma$ can be approximated as
\be
\Gamma(r,\tau,v=0) \sim D''(r) \tau^2 /2  .
\label{eq:Gamma.zero}
\ee
Hence we see that for small lags the rms error in $\hat{D}(\tau)$
varies as $\tau^2$, for any well-behaved $D(\tau )$ function.

To make further progress, we assume a particular form for
$D(\tau)$.  The cases of interest are where $D$ is a power law
$\propto \tau^a$  for small lags, and we adopt the following
simple model (with $1 \simless a \simless 2$)
\be
D(\tau) = D_{\infty} \tau^a / (\tau_o^a + \tau^a)  .
\label{eq:Dpowerlaw}
\ee
We can use this in equation(\ref{eq:vard}) for values of $r \simgreat \tau$,
but the integral is not done simply.  We have investigated the integral
numerically and find the following to be a useful
approximation which connects the behavior for
$\tau \simless\tau_o$ and $\tau \le \tau_o$.
\be
\hat{D}(\tau )_{\rm rms} \sim   D_{\infty} 2 \sqrt{\frac{\tau_o}{T}}
\frac{\tau^2}{\tau_o^2 + \tau^2}  .
\label{eq:drms}
\ee
The accuracy of this approximation is best judged from the simulations
described below.  The important point about this analysis is that
for exponents $a$ less than 2,  as $\tau$ decreases the error in
$\hat{D(\tau )}$ decreases more quickly than $D(\tau )$ itself,
and so the fractional error in $\hat{D}$ decreases with small $\tau$, as
illustrated in the simulations of Coles and Harmon (1989).

The second limiting condition to be discussed is when the duration $T$
is short compared to $\tau_o$.  Then the triangular weight in
equation(\ref{eq:cov-uv}) must be included. For the variance
in $\hat{D}$ we again set $v=0$ and since $T \ll \tau_o$ we can approximate
$D(\tau) \sim D_{\infty} (\tau/\tau_o)^a$.  Approximating the integral
for $2 > a > 1.5$ gives
\be
\hat{D}(\tau )_{\rm rms} \sim   D_{\infty} (\tau/T')^2
(T'/\tau_o)^{2a} a .
\ee
This result when normalized by $D(\tau)$ shows a fractional error
varying $\propto (\tau/T')^{2-a}$, again decreasing with small $\tau$.

Eq.(\ref{eq:cov-uv}) also allows us to consider the covariance
of $D$ estimates at lags $\tau_1$ and $\tau_2$.  Examining
the influence of non-zero $v$, equation(\ref{eq:Gammdef})
shows that as $v$ increases from zero it only has a substantial influence when
it is comparable to $u$.  In other words the estimates are only independent
when the difference in time lags is comparable to their mean.
A consequence of this correlation between estimates of $D$
is that one can estimate the exponent $a$ with reasonable accuracy
even with small time lags and data duration less than $\tau_o$.

To investigate the validity of the foregoing analysis and its approximations
we have simulated random time series with a power law structure function,
applied the structure function estimator and studied its error properties.
The process was specified by its power spectrum which was
$P(f) = P_o [(1+(f/f_o)^2]^{-(a+1)/2}$.  This has a high frequency form
$\propto f^{-a-1}$, and the corresponding structure function
is $\propto \tau^{a}$ for $\tau f_o \ll 1$.


\begin{figure}[tbh]
\psfig{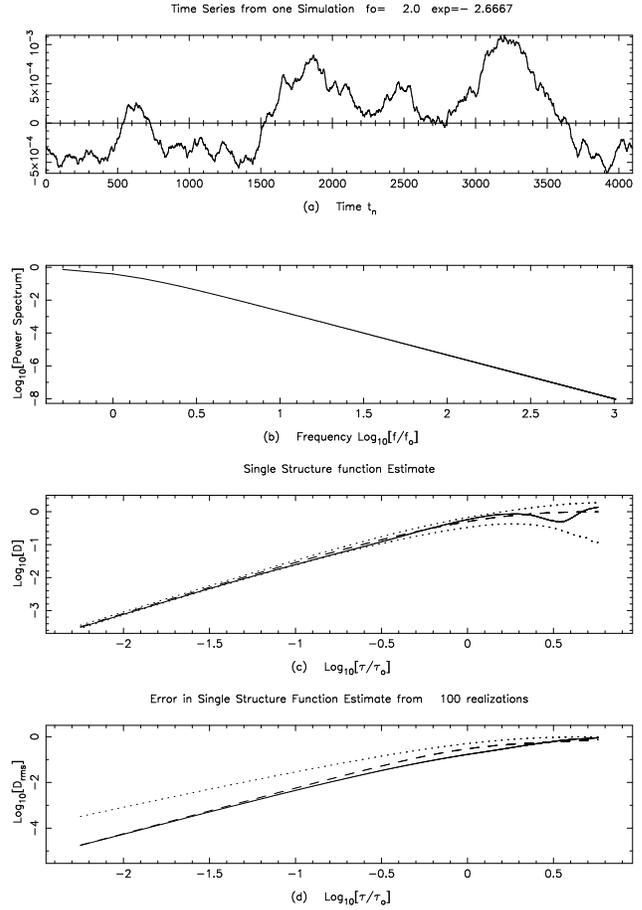}
\caption{Simulation of a time series with a power law structure function
($\propto t^{5/3}$).
(a) Single sample of the time series
(b) Ensemble average power spectrum.
(c) Solid line is an estimate of the structure function
from the single realization shown in (a);  dashed line is the
theoretical average structure function and dotted lines are
plus and minus the standard deviation in the structure
function estimate from (d).
(d) Cumulated estimate of the rms error in the structure function estimates
from 100 realizations (solid line); dashed line is the analytical
approximation to the error given by equation (B12)
dotted line shows the average structure function on the same scale,
illustrating that the relative rms error in the estimate improves at
small time lags.}
\label{fig:dsim}
\end{figure}


\begin{figure}[tbh]
\psfig{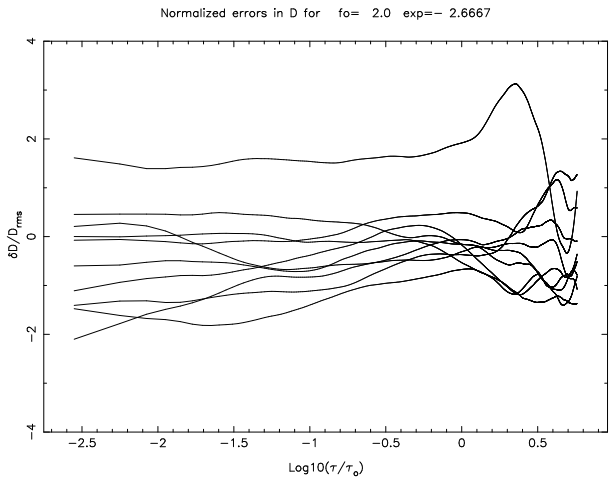}
\caption{Simulation of a time series with a power law structure function
($\propto t^{5/3}$).
Set of 10 errors $\hat{D}(\tau ) - D(\tau )$ normalized by the
expected error from equation (B12).
This shows that for $\tau$ less than about 0.3 $\tau_o$
the estimation errors are highly correlated, and that
for lags above $\tau_o$ the errors are correlated over a range
in lag of about $\tau_o$.}
\label{fig:derr}
\end{figure}

We use the standard method
of simulating the process in the frequency domain. Independent
zero mean Gaussian random variables of variance  $P(f)/2$
are stored into the real and imaginary parts of the complex
Fourier representation of the series. After making it an Hermitian
spectrum, an FFT is used to obtain the (real) random variable $X_n$
at time samples $t_n$. Figure \ref{fig:dsim}(a) shows 
a single example with $a=5/3$
and $0 < n < 4095$.  The turnover frequency is set low, to give
over two decades with power law behavior; this correspond to
sample interval of $\tau_o/180$ and $T = 23 \tau_o$.  We applied
the structure function estimator eq (\ref{eq:dhat,disc})
to the time series, and repeated the process on each new realization
of the time series.  This was repeated to obtain many independent
estimates $\hat{D}(\tau )$, whose mean and rms were computed versus $\tau$.
In Figure \ref{fig:dsim}(b),
we show the ensemble average power spectrum; in Fig \ref{fig:dsim}(c) we show
as a heavy line an estimation $\hat{D}$
from the single time series of Fig \ref{fig:dsim}(a);
the dashed line is the theoretical average $D(\tau )$; the dotted lines
show this mean plus and minus two standard deviations derived from 50
realizations of $\hat{D}$.  The fourth panel
Fig \ref{fig:dsim}(d) shows the cumulated
estimates of the rms error in $\hat{D}(\tau )$ as a solid line
and the analytical approximation of equation(\ref{eq:drms}) as a dashed line,
which evidently provides a very useful approximation to the error.
The dotted line shows $D(\tau)$ with the logarithmic scales it illustrates
how the fractional error in $\hat{D}$ becomes smaller for small
$\tau$.  The structure function estimates are plotted from
2 to 2048 sample intervals.

The degree of correlation between the $\hat{D}$ estimates at two
time lags, can be found as a function of the two lags.  Instead of
displaying such a correlation surface, we simply plot in Figure
\ref{fig:derr} a set of 10 errors $\hat{D}(\tau ) - D(\tau )$
normalized by the expected error from equation(\ref{eq:drms}).
This shows that for $\tau$ less than about 0.3 $\tau_o$
the estimation errors are highly correlated, and that
for lags above $\tau_o$ the errors are correlated over a range
in lag of about $\tau_o$.

\section{Scattering Measure Estimates}
\label{sec:sm}

For a screen of thickness $\Delta$ at distance $xL$
equations (\ref{eq:s_scr}) together give the relation
of $m_{\rm w}^2$ to $SM$. With $\alpha = 5/3$ this gives
\bea
SM = C_N^2 \Delta = 0.01465    \times \nonumber \\ 
m_{\rm w}^2 [x(1-x)L \lambda /2\pi ]^{-5/6} r_e^{-2} \lambda^{-2}
\label{eq:sm.mwscr}
\eea
For our measurements this gives
$SM \sim 2.8 \times 10^{-6}$m$^{-20/3}$kpc, for a screen
at 234 pc and (by coincidence) the same value for a screen at
72 pc (the edge of the bubble in the Bhat et al\ model).
If, instead, we assume a uniform scattering medium, then
\be
SM = 0.0664 \;m_{\rm w}^2 (L\lambda/2\pi)^{-5/6} r_e^{-2} \lambda^{-2}
\label{eq:sm.mwext}
\ee
and we obtain $SM \sim 3.0 \times 10^{-6}$m$^{-20/3}$kpc
for our pulsar measuremnt.

$SM$ estimates were also been found by Cordes (1986),
in which he measured the diffractive decorrelation
bandwidth $\delta \nu_d$ for various pulsars
and converted to $C_N^2$ for an equivalent
uniform scattering medium. However, the formula used, his equation (6),
derived by Cordes, Weisberg and Boriakoff (1993) [CWB]
differs from those in the more recent model by TC93.
For a spherical wave point source in a uniform
scattering medium at distance $L$ from the observer,
the relation can be expressed in the general form
based on CWB equation (6):
\be
SM = A \, \nu^{\alpha+2} L^{-\alpha/2} \delta \nu_d^{-\alpha/2} .
\label{eq:sm.deltanu}
\ee
CWB derived $A$ in their Appendix A and then increased
it by a factor 6 to account for the effective
line-of-sight weighting.
Cordes (1986) used the resulting expression in deriving an
average $C_N^2 = SM/L$ for ISS measurements of over
70 pulsars.  However, in constructing their model
for the distribution of dispersing and scattering
electrons in the ISM, TC93 used a different
formula (their equation 8).  It is expressed in terms of
the scatter broadening time, which they relate by
$\tau_s = (2 \pi \delta \nu_d)^{-1}$ to the decorrelation bandwidth.
Putting this into the same format as above, their formula
corresponds to $A = 6.3 \times 10^{-4}$ where $SM$ is in the
units m$^{-20/3}$kpc, $\nu$ is in GHz, $L$ is in kpc
and $\delta \nu_d$ is in MHz.  We note that this is
about one third of the value ($A = 0.002$) used by CWB and
Cordes (1986).  In the study of scattering in the local bubble
Bhat et al.\ (1998) used this larger value also.

The work of Lambert and Rickett (1999) allows us to obtain an
independent theoretical evaluation for $A$.  They computed
the spherical wave two-frequency ``diffractive'' correlation function
for three spectral models, including the Kolmogorov spectrum.
Asymptotically strong scattering
was assumed in a uniform scattering medium, leading
to computed intensity correlation functions, from which precise
values of the decorrelation bandwidth were obtained. These are given
in terms of a normalized decorrelation width $v_d$ for each spectrum
model; the value $v_d = 2.320$ was found for the Kolomogorov spectrum.
The decorrelation bandwidth is then
$\delta \nu_d = \nu v_d s_{0s}^2/(L\lambda/2\pi)$.
Here $s_{0s}$ is the observed spherical wave
field coherence scale, where
the phase structure function equals one.
It is related by standard formulae to $SM$ (e.g.\ their
equations 11 and 13). Eliminating $s_{0s}$ we find:
\be
A_{\rm ext} = \frac{ (2\pi v_d)^{\alpha/2} \Gamma(1+\alpha/2) \alpha
(\alpha+1) 2^{\alpha} }
{ 8 \pi^2 \Gamma(1-\alpha/2) r_e^2 c^{2+\alpha/2} } ,
\ee
where $c$ is the speed of light.
With $\alpha = 5/3$ and using the same units as above, we
find $A = 6.25 \times 10^{-4}$ in very close agreement with
the formula used by TC93, and 0.315 times
the value used in work following Cordes (1986) and CWB.
When comparing observations and theory
with an accuracy of better than a factor of 3,
observers should use the constant $A = 6.3 \times 10^{-4}$
in equation (\ref{eq:sm.deltanu}).  In particular,
in the models of Bhat et al.\ $C_N^2$ should be
reduced by a factor of three over their published values.
We have made the appropriate changes in listing SM
values for pulsar B0809+74 in Table 1.
For a screen at distance $xL$ from the Earth,
a similar analysis leads to:
\be
A_{\rm scr} = 0.179 A_{\rm ext} [x(1-x)]^{-5/6}
\ee



\clearpage

\begin{thebibliography}{}

\bibitem[ ]{}
  Backer, D. C. 1975, \apj, 190, 395

\bibitem[ ]{}
  Bhat, N. D. Ramesh, Gupta, Y. \& Rao, A. Pramesh 1998,
 \apj, 500, 262

\bibitem[BGR 1999a]{BRG99a}
  Bhat, N. D. Ramesh., Rao, A. P., \&  Gupta, Y. 1999a,
  \apj, 514, 249

\bibitem[BGR 1999b]{BRG99b}
  Bhat, N. D. Ramesh, Gupta, Y. \& Rao, A. Pramesh 1999b,
 \apj, 514, 272

\bibitem[Britton et al.\ 1998]{Britton98}
  Britton, M. C., Gwinn, C. R. \& Ojeda, M. J. 1998, \apj, 501, L101

\bibitem[ ]{}
  Breitschwerdt, D., Freyberg, M. J. and Tr\"umper, J. 1998,
``The Local Bubble and Beyond'', Proceedings of IAU Colloquium 166,
 Ed. Breitschwerdt, D., Freyberg, M. J. and Tr\"umper, J., p1


\bibitem[ ]{}
  Coles, W. A. \& Harmon, J. K. 1989. \apj, 227, 1023

\bibitem[ ]{}
  Coles, W. A., Frehlich, R. G., Rickett, B. J. \& Codona, J. L. 1987.
\apj, 315, 666

\bibitem[ ]{}
  Cordes, J. M. 1986, \apj, 311, 183

\bibitem[ ]{}
  Cordes, J. M., Weisberg, J. M. \& Boriakoff, V. 1985, \apj, 288, 221 [CWB]

\bibitem[ ]{}
  Cordes, J. M. \& Rickett, B. J. 1998,  \apj, 507, 846

\bibitem[ ]{}
  G\'{e}nova, R., Beckman, J. E. \& Alamo, J. R. 1998, in
  ``The Local Bubble and Beyond'', Proceedings of IAU Colloquium 166,
   Ed. Breitschwerdt, D., Freyberg, M. J. \& Tr\"umper, J., p 195

\bibitem[ ]{}
  Gould, D. M., Lyne, A. G., 1998, \mnras, 301, 235


\bibitem[Gupta et al.\ 1994]{Gupta94}
  Gupta, Y., Rickett, B. J., \& Lyne, A. G. 1994,
  \mnras, 269, 1035

\bibitem[ ]{}
  Jenkins, G. M. \& Watts, D. G. 1969, ``Spectral Analysis and its
  Applications'' (San Francisco: Holden-Day)

\bibitem[ ]{}
  Lallement, R. 1998, in
  ``The Local Bubble and Beyond'', Proceedings of IAU Colloquium 166,
   Ed. Breitschwerdt, D., Freyberg, M. J. \& Tr\"umper, J., p19

\bibitem[ ]{}
 Lambert, H. C. \& Rickett, B. J. 1999, \apj, 517, 299

\bibitem[ ]{}
  Lyne, A. G., Anderson, B. \& Salter, C. J. 1982, \mnras, 213, 613

\bibitem[ ]{}
  Malofeev, V. M., Shishov, V. I., Sieber, W., Jessner, A.,
  Kramer, M. \& Wielebinski, R., 1996, \aap, 308, 180

\bibitem[ ]{}
  Narayan, R. 1992, Phil. Trans. R. Soc. London A, 341, 151

\bibitem[ ]{}
  Otterbein, K., Krichbaum, T. P., Kraus, A., Lobanov, A. P., 
  Witzel, A., Wagner, S. J.,  \& Zensus, J. A. 1998, \aap,
  334, 489

\bibitem[ ]{}
  Phillips, J. A. \& Clegg, A. W. 1992, Nature, 360, 137

\bibitem[ ]{}
  Quirrenbach, A., Witzel, A., Krichbaum, T. P.,
  Hummel, C. A., Wegner, R.,  Schalinski, C. J., Ott, M.,
  Alberdi, A., \& Rioja, M. 1992, 258, 279

\bibitem[ ]{}
  Prokhorov, A. M., Bunkin, F. V., Gochelashvily \& K. S., Shishov, V. I.
1975.
  Proc. I.E.E.E. 63, 790

\bibitem[ ]{}
  Rickett, B. J. 1970,  \mnras, 150, 67

\bibitem[ ]{}
  Rickett, B. J. 1990,  \araa, 28, 561

\bibitem[ ]{}
   Romani, R. W., Narayan, R., \& Blandford, R. 1986,
  \mnras, 220, 19

\bibitem[ ]{}
  Snowden, S. L. 1998, in
  ``The Local Bubble and Beyond'', Proceedings of IAU Colloquium 166,
   Ed. Breitschwerdt, D., Freyberg, M. J. \& Tr\"umper, J., p103

\bibitem[]{}
  Spangler, S. R., \& Gwinn, C. R. 1990, \apj, 353, L29

\bibitem[ ]{}
  Tatarskii, V. I. \& Zavorotnyi, V. U. 1980, Progress in
  Optics, 18, 207

\bibitem[ ]{}
  Taylor, J. H. \& Cordes, J. M. 1993,  \apj, 411, 674

\bibitem[ ]{}
  Taylor, J. H., Manchester, R. N., \& Lyne A. G. 1993,
  \apjs, 88, 529

\bibitem[ ]{}
  Wagner, S. J., Witzel, A., Heidt, J., Krichbaum, T. P.,
  Qian, S. J., Quirrenbach, A., Wegner, R., Aller, H., Aller, M.,  
  Anton, K., Appenzeller, I., Eckart, A., Kraus, A., Naundorf, C.,
  Kneer, R., Steffen, W., \& Zensus, J. A. 1996,  \aj, 111, 2187

\end{thebibliography}
\end{document}